\documentclass[12pt,dvips]{article}

\usepackage{amsmath,epsfig,ulem}
\usepackage{rotating}
\usepackage{hhline}
\usepackage{array}
\usepackage{amssymb}
\usepackage{color}

\begin{document}
\tolerance=100000

\newcommand{\imag}{\Im {\rm m}}
\newcommand{\real}{\Re {\rm e}}

\def\tablename{\bf Table}%
\def\figurename{\bf Figure}%

\newcommand{\sts}{\scriptstyle}
\newcommand{\ngs}{\!\!\!\!\!\!}
\newcommand{\rb}[2]{\raisebox{#1}[-#1]{#2}}
\newcommand{\CP}{${\cal CP}$~}
\newcommand{\sbomu}{\frac{\sin 2 \beta}{2 \mu}}
\newcommand{\kmol}{\frac{\kappa \mu}{\lambda}}
\newcommand{\s}{\\ \vspace*{-3.5mm}}
\newcommand{\lsim}{\raisebox{-0.13cm}{~\shortstack{$<$\\[-0.07cm] $\sim$}}~}
\newcommand{\gsim}{\raisebox{-0.13cm}{~\shortstack{$>$\\[-0.07cm] $\sim$}}~}
\newcommand{\kr}{\color{red}}
\newcommand{\RF}{${\mathbb{R} \!\!\! / \,\,\,}$}
\newcommand{\Rp}{${R_p \!\!\!\!\!\! / \,\,\,\,\,}$}

\begin{titlepage}

\begin{flushright}
IFT 07--007\\
\today
\end{flushright}

\vskip 1.5cm

\begin{center}
{\large \bf CP violation at one loop in the polarization-independent
chargino production in $e^+e^-$ collisions
}\\[0.9cm]
{\normalsize K. Rolbiecki and  J. Kalinowski}\\[0.8cm]
{\it  Institute of Theoretical Physics, University of Warsaw\\
Ho\.za 69, PL--00681 Warsaw,
          Poland}
\end{center}

\renewcommand{\thefootnote}{\fnsymbol{footnote}}
\vspace{2.3cm}

\begin{abstract}
\noindent Recently Osland and Vereshagin noticed, based on sample
calculations of some box diagrams, that in unpolarised $e^+e^-$
collisions CP-odd effects in the non-diagonal chargino-pair
production process are generated at one-loop. Here we perform a full
one-loop analysis of these effects and point out that in some cases
the neglected vertex and self-energy contributions may play a
dominant role. We also show that CP asymmetries in chargino
production are sensitive not only to the phase of $\mu$ parameter in
the chargino sector but also to the phase of stop trilinear coupling
$A_t$.
\end{abstract}

\end{titlepage}

\newpage
\renewcommand{\thefootnote}{\fnsymbol{footnote}}

\section{Introduction}
The electroweak sector of the Standard Model (SM) contains only one
CP-violating phase which arises in the Cabibbo-Kobayashi-Maskawa
(CKM) quark mixing matrix. Adding right-handed neutrinos to account
for non-zero neutrino masses and their mixing opens up a possibility
of new CP-violating phases in the Maki-Nakagawa-Sakata (MNS) lepton
mixing matrix. While the observed amount of CP violation in the $K$
and $B$ can be accommodated within the SM, another (indirect) piece
of evidence of CP violation, the baryon asymmetry in the universe,
requires a new source of CP violation~\cite{genesis}. Thus new
CP-violating phases must exist in nature.

Supersymmetric extensions of the SM  introduce a plethora of CP
phases in soft supersymmetry breaking terms. This poses a SUSY CP
problem, since if the phases are large ${\cal O}(1)$, SUSY
contributions to the lepton and neutron  EDMs can be too large to
satisfy current experimental constraints \cite{susycp}. Many models
have been proposed \cite{overcome} to overcome this problem: fine
tune phases to be small, push sparticle spectra (especially squarks
and sleptons) above a TeV scale to suppress effects of large phases
on the EDM, arrange for internal cancelations etc.

In the absence of any reliable theory that forces in a natural way
the phases to be vanishing or small, it is mandatory to consider
scenarios with some of the phases large and arranged consistent with
experimental EDM data. In such CP-violating scenarios charginos and
neutralinos (denoted generically by $\tilde{\chi}$) might be light enough to be
produced at $e^+e^-$ colliders, and many phenomena will be affected
by non-vanishing phases: sparticle masses, their decay rates and
production cross sections, SUSY contributions to SM processes etc.
However, most unambiguous way to study the presence of CP-violating
phases would be in some CP-odd observables measurable at future
accelerators.

To build a CP-odd observable in a two-fermion $\to$ two-fermion
process, e.g.\ $e^+e^-\to\tilde{\chi}_i\tilde{\chi}_j$, typically
one uses spin information of one of the particles involved. For
example, a measurement of the fermion polarization $s$ transverse to
the production plane~\cite{Kizukuri:1993vh} allows to build a CP-odd
observable $s\cdot(p_e\times p_{\tilde{\chi}})$. This requires
either transverse beam polarization and/or spin-analyser of produced
$\tilde{\chi}$'s via angular distributions of their decay
products~\cite{spin}. Another possibility is to look into triple
products involving momenta of the decay products of charginos in
case of longitudinal polarization of the beams~\cite{Kittel:2004kd}.

However, CP-odd effects can also be detected in simple
event-counting experiments if several processes are measured. One
example is provided by non-diagonal neutralino-pair production in
$e^+e^-$ annihilation with unpolarized beams: observing the
$\tilde{\chi}^0_i\tilde{\chi}^0_j$,
$\tilde{\chi}^0_i\tilde{\chi}^0_k$ and
$\tilde{\chi}^0_j\tilde{\chi}^0_k$ pairs to be excited {\it all} in
S-wave near respective thresholds signals CP-violation in the
neutralino sector at tree-level~\cite{Choi:2001ww}. Alternatively,
unambiguous evidence for CP-violation in the neutralino system is
provided by the observation of simultaneous sharp S–-wave
excitations of both the production of any non–-diagonal neutralino
pair $\tilde{\chi}^0_i\tilde{\chi}^0_j$ near threshold and the
$f\bar f$ invariant mass distribution of the decay
$\tilde{\chi}^0_j\to \tilde{\chi}^0_i f \bar f$ near the end
point~\cite{Choi:2003hm}.

Recently Osland and Vereshagin pointed out that in the non-diagonal
chargino pair production process
$e^+e^-\to\tilde{\chi}_1^\pm\tilde{\chi}_2^\mp$ a CP-odd observable
can be constructed from unpolarized cross sections at
one-loop~\cite{Osland:2007xw}. Their simplified  numerical analysis
based on  only some of the box diagrams shows that, indeed, the
CP-violation induced by the complex higgsino mass parameter $\mu$
may in principle be observed in this reaction without any spin
detection and with unpolarized initial beams.

In this note we perform a full one-loop analysis of the non-diagonal
chargino pair production.  First we recapitulate the MSSM chargino
sector at tree level and show explicitly that such a CP asymmetry
vanishes. In Section 3 we discuss the CP asymmetry at one loop. We
note that a non-zero asymmetry requires not only complex couplings
but also absorptive parts of Feynman diagrams. In  Section 4 we
present numerical results for the CP asymmetry and discuss  relative
weights of various contributions. We consider  effects of both the
complex higgsino mass parameter and the complex trilinear scalar
coupling in the top squark sector. Section 5 summarizes and
concludes our analysis.

\section{MSSM chargino sector at tree level} \label{sec2}
In the minimal supersymmetric extension of the Standard Model
(MSSM), the tree-level mass matrix of the  spin-1/2 partners of the
charged gauge and Higgs bosons, $\tilde{W}^-$ and $\tilde{H}^-$,
takes the form
\begin{eqnarray}
{\cal M}_C=\left(\begin{array}{cc}
  M_2       &   \sqrt{2}  m_W\cos\beta \\[2mm]
   \sqrt{2} m_W \sin\beta &     \mu
                  \end{array}\right)\, ,
\label{eq:massmatrix}
\end{eqnarray}
where $M_2$ is the SU(2) gaugino mass, $\mu$ is the higgsino mass
parameter, and $\tan\beta$ is the ratio $v_2/v_1$ of the vacuum
expectation values of the two neutral Higgs fields. By
reparametrization of the fields, $M_2$ can be taken real and
positive, while $\mu$ can be complex $\mu=|\mu|\,\,{\rm
e}^{i\Phi_\mu}$. Since the chargino mass matrix ${\cal M}_C$ is not
symmetric, two different unitary matrices acting on the left- and
right-chiral $(\tilde{W},\tilde{H})_{L,R}$ two-component states
\begin{eqnarray}
U_{L,R}\left(\begin{array}{c}
             \tilde{W}^- \\
             \tilde{H}^-
             \end{array}\right)_{L,R} =
       \left(\begin{array}{c}
             \tilde{\chi}^-_1 \\
             \tilde{\chi}^-_2
             \end{array}\right)_{L,R}
\end{eqnarray}
are needed to diagonalize it. The unitary matrices $U_L$ and $U_R$
can be parameterized in the following way \cite{Choi:charg}
($s_\alpha=\sin\alpha$, $c_\alpha=\cos\alpha$):
\begin{eqnarray}
&&
U_L=\left(\begin{array}{cc}
             c_{\phi_L} & {\rm e}^{-i\beta_L}s_{\phi_L} \\
            -{\rm e}^{i\beta_L}s_{\phi_L} & c_{\phi_L}
             \end{array}\right), \qquad
U_R=\left(\begin{array}{cc}
             {\rm e}^{i\gamma_1} & 0 \\
             0 & {\rm e}^{i\gamma_2}
             \end{array}\right)
        \left(\begin{array}{cc}
             c_{\phi_R} & {\rm e}^{-i\beta_R}s_{\phi_R} \\
            -{\rm e}^{i\beta_R}s_{\phi_R} & c_{\phi_R}\,
             \end{array}\right)\, .
\end{eqnarray}
As far as $\Phi_\mu$ dependence is concerned, the mass eigenvalues
and rotation angles $\cos2\phi_{L,R}$, $\sin2\phi_{L,R}$, being
CP-even,  are functions of $\cos\Phi_{\mu}$ only. On the other hand,
the four phases $\beta_{L,R}$ and $\gamma_{1,2}$ are CP-odd since
their tangents depend linearly on $\sin\Phi_\mu$; all four phases
vanish in CP-invariant case for which $\Phi_\mu=0$ or $\pi$.

Charginos can copiously be produced at prospective $e^+e^-$ linear
colliders~\cite{futurecoll}. At tree-level they are produced  via
the $s$-channel $\gamma,Z$ exchange and $t$-channel electron
sneutrino exchange. Photon exchange contributes only to the
production of diagonal pairs $\tilde{\chi}_1^+ \tilde{\chi}_1^-$ and
$\tilde{\chi}_2^+ \tilde{\chi}_2^-$. The production amplitude, after
a Fierz transformation of the $t$-channel contribution,
\begin{eqnarray}
{\cal A}[e^+e^-\rightarrow\tilde{\chi}^-_i\tilde{\chi}^+_j]
  =\frac{e^2}{s}Q^{ij}_{\alpha\beta}
   \left[\bar{v}(e^+)\gamma_\mu P_\alpha  u(e^-)\right]
   \left[\bar{u}(\tilde{\chi}^-_i) \gamma^\mu P_\beta
               v(\tilde{\chi}^+_j) \right],
\label{eq:production amplitude}
\end{eqnarray}
is expressed in terms of four bilinear charges
$Q^{ij}_{\alpha\beta}$, defined by the chiralities
$\alpha,\beta=L,R$ of the lepton and chargino currents. The charges
take the form
\begin{eqnarray}
&& Q^{ij}_{RL}=\delta_{ij} D_R + C^L_{ij} F_R,
          \qquad
   Q^{ij}_{LL}=\delta_{ij} D_L  + C^L_{ij} F_L, \nonumber \\
&& Q^{ij}_{RR}=\delta_{ij} D_R +C^R_{ij} F_R,
    \qquad
   Q^{ij}_{LR}=\delta_{ij} D_L+ C^R_{ij} F_L+
   {\textstyle\frac{D_{\tilde{\nu}}}{4s^2_W}} (\delta_{ij}-
        C^R_{ij})\, ,
\label{eq:[12]}
\end{eqnarray}
with $s$-, $t$-channel propagators
$   D_L=1+{\textstyle \frac{D_Z}{s_W^2 c_W^2}}
      (s_W^2 -{\textstyle\frac{1}{2}})(s_W^2-{\textstyle\frac{3}{4}})$,
$   F_L={\textstyle\frac{D_Z}{4s_W^2 c_W^2}}
      (s^2_W-{\textstyle\frac{1}{2}})$,
$   D_R=1+\frac{D_Z}{c_W^2}(s_W^2-\frac{3}{4})$, $
F_R=\frac{D_Z}{4c_W^2}$, $D_Z=s/(s-m^2_Z)$,
$D_{\tilde{\nu}}=s/(t-m^2_{\tilde{\nu}})$; the $\tilde{\nu}_e$
exchange contributes only to the $LR$ amplitude. The coefficients
$C^L_{ij}$ are functions of $U_L$ as follows
\begin{eqnarray}
C^L_{11}=-\cos2\phi_{L}, \quad C^L_{22}=\cos2\phi_{L}, \quad
C^L_{12}=e^{-i\beta_L} \sin2\phi_{L}, \quad C^L_{21}=e^{i\beta_L}
\sin2\phi_{L}\, , \label{cij_coeff}
\end{eqnarray}
for $C^R$ replace $\phi_L\to\phi_R$ and
$\beta_L\to\beta_R-\gamma_1+\gamma_2$.

Note that the phases $\beta_L,\beta_R,\gamma_1,\gamma_2$ enter only
non-diagonal $\{12\}$ and $\{21\}$ amplitudes.  However, after
summing over chargino helicities, the dependence on these phases
disappears in the polarized differential cross section for the
$e^+e^-\to\tilde{\chi}^-_i \tilde{\chi}^+_j$. Defining the polar
angle $\theta$ and the azimuthal angle $\phi$ of $\tilde{\chi}^-_i$
with respect to the $e^-$ momentum direction and the $e^-$
transverse polarization vector, respectively, the polarized
differential cross section is given by~\cite{Choi:charg}
\begin{eqnarray}
&& {\rm d}\sigma^{ij}\equiv \frac{{\rm d}\sigma^{\{ij\}}}{{\rm d}\cos\theta \,{\rm d}\phi}
  =\frac{\alpha^2}{16\, s}\, \lambda^{1/2} \bigg[
     (1-P_L\bar{P}_L)\,\Sigma_{\rm unp}+(P_L-\bar{P}_L)\,\Sigma_L
  +P_T\bar{P}_T\cos(2\phi-\eta)\,\Sigma_T \bigg]\, \label{eq:diffx}
\end{eqnarray}
where $P$=$(P_T,0,P_L)$ [$\bar{P}$=$(\bar{P}_T
\cos\eta,\bar{P}_T\sin\eta, -\bar{P}_L)$] is the electron [positron]
polarization vector;
$\lambda=[1-(\mu_i+\mu_j)^2][1-(\mu_i-\mu_j)^2]$ with
$\mu_i=m_i/\sqrt{s}$. The distributions $\Sigma_{\rm unp}$,
$\Sigma_L$ and $\Sigma_T$ depend only on the polar angle $\theta$
and can be expressed as (the superscripts $\{ij\}$ labeling the
produced chargino pair are understood)
\begin{eqnarray}
\Sigma_{\rm unp}&=& 4\,\{[1-(\mu^2_i - \mu^2_j)^2
                   +\lambda\cos^2\theta]Q_1
                   +4\mu_i\mu_j Q_2+2\lambda^{1/2} Q_3\cos\theta\},
                  \nonumber\\
\Sigma_{LL}     &=& 4\,\{[1-(\mu^2_i - \mu^2_j)^2
                   +\lambda\cos^2\theta]Q'_1
                   +4\mu_i\mu_j Q'_2+2\lambda^{1/2} Q'_3\cos\theta\},
                  \nonumber\\
\Sigma_{TT}     &=&-4\lambda \sin^2\theta\,\, Q_5.
\end{eqnarray}
The eight quartic charges for each of the production processes of
the diagonal and mixed chargino pairs, expressed in terms of
bilinear charges, are collected in Table~1, including the
transformation properties under P and CP.\\
\begin{table*}[\hbt]
\caption[{\bf Table 1:}]{\label{tab:quartic}
{\it The quartic charges of the chargino system.}}
\begin{center}
\begin{tabular}{|c|c|l|}\hline
 &  &  \\[-4mm]
${\rm P}$ & ${\rm CP}$ & { }\hskip 2cm Quartic charges \\\hline \hline
 even    &  even     & $Q_1 =\frac{1}{4}\left[|Q_{RR}|^2+|Q_{LL}|^2
                       +|Q_{RL}|^2+|Q_{LR}|^2\right]$ \\[2mm]
         &           & $Q_2 = \frac{1}{2}{\rm Re}\left[Q_{RR}Q^*_{RL}
                       +Q_{LL}Q^*_{LR}\right]$ \\[2mm]
         &           & $Q_3 = \frac{1}{4}\left[|Q_{RR}|^2+|Q_{LL}|^2
                       -|Q_{RL}|^2-|Q_{LR}|^2\right]$ \\[1mm]
         &           & $Q_5=\frac{1}{2}{\rm Re} \left[Q_{LR}Q^*_{RR}
                       +Q_{LL}Q^*_{RL}\right]$ \\[1mm]
\cline{2-3}
         &  odd      & $Q_4=\frac{1}{2}{\rm Im}\left[Q_{RR}Q^*_{RL}
                       +Q_{LL}Q^*_{LR}\right]$\\[2mm] \hline \hline
 odd     &  even     & $Q'_1=\frac{1}{4}\left[|Q_{RR}|^2+|Q_{RL}|^2
                        -|Q_{LR}|^2-|Q_{LL}|^2\right]$\\[1mm]
         &           & $Q'_2=\frac{1}{2}{\rm Re}\left[Q_{RR}Q^*_{RL}
                        -Q_{LL}Q^*_{LR}\right]$ \\[2mm]
         &           & $Q'_3=\frac{1}{4}\left[|Q_{RR}|^2+|Q_{LR}|^2
                        -|Q_{RL}|^2-|Q_{LL}|^2\right]$\\[1mm]
\hline
\end{tabular}
\end{center}
\end{table*}\\[-2mm]
The charges $Q_1$ to $Q_5$ are manifestly parity-even, $Q'_1$ to
$Q'_3$ are parity-odd. The charges $Q_1$ to $Q_3$, $Q_5$, and $Q'_1$
to $Q'_3$ are CP-invariant. Only $Q_4$ changes sign under CP
transformations.

From above expressions it is evident that even for transverse beam
polarization the differential cross section $\sim\Sigma_{TT}$ is
CP-even. Also the differential distributions for non-diagonal
chargino pairs $\tilde{\chi}^-_1\tilde{\chi}^+_2$ and $\tilde{\chi}^-_2\tilde{\chi}^+_1$ are equal,
so the asymmetry
\begin{eqnarray}
&& A_{12}=\frac{\int_{-1}^1({\rm d}\sigma^{\{12\}}-
  {\rm d}\sigma^{\{21\}}) {\rm d} \cos\theta}{\int_{-1}^1({\rm d}\sigma^{\{12\}}+
  {\rm d}\sigma^{\{21\}}){\rm d} \cos\theta}
  \label{CPasy}
\end{eqnarray}
at tree level vanishes in CP-noninvariant theories. The CP-odd
quartic charge $Q_4$ can only be probed by observables sensitive to
the chargino polarization component normal to the production plane
in mixed $e^+e^-\to \tilde{\chi}^\pm_1\tilde{\chi}^\mp_2$
processes~\cite{Choi:charg}. Thus, at tree-level one cannot build a
CP-odd observable from chargino polarized cross sections alone.

Due to Poincar\'{e} invariance the unpolarized differential cross
section $\sim\Sigma_{\rm unp}$ may depend only on masses $m_i, m_j$
and on two independent scalar variables $s$ and $t$. As a result,
the unpolarized differential cross-sections  for equal-mass fermions
$m_i=m_j$ in the final state are always CP-even. However, if the
chargino species are different, beyond tree level the CP-violating
terms can arise  even in the unpolarized
cross-section~\cite{Osland:2007xw}.

\section{CP-odd asymmetry at one-loop}

\begin{figure}[!t]
 \begin{center}
\includegraphics[scale=0.9]{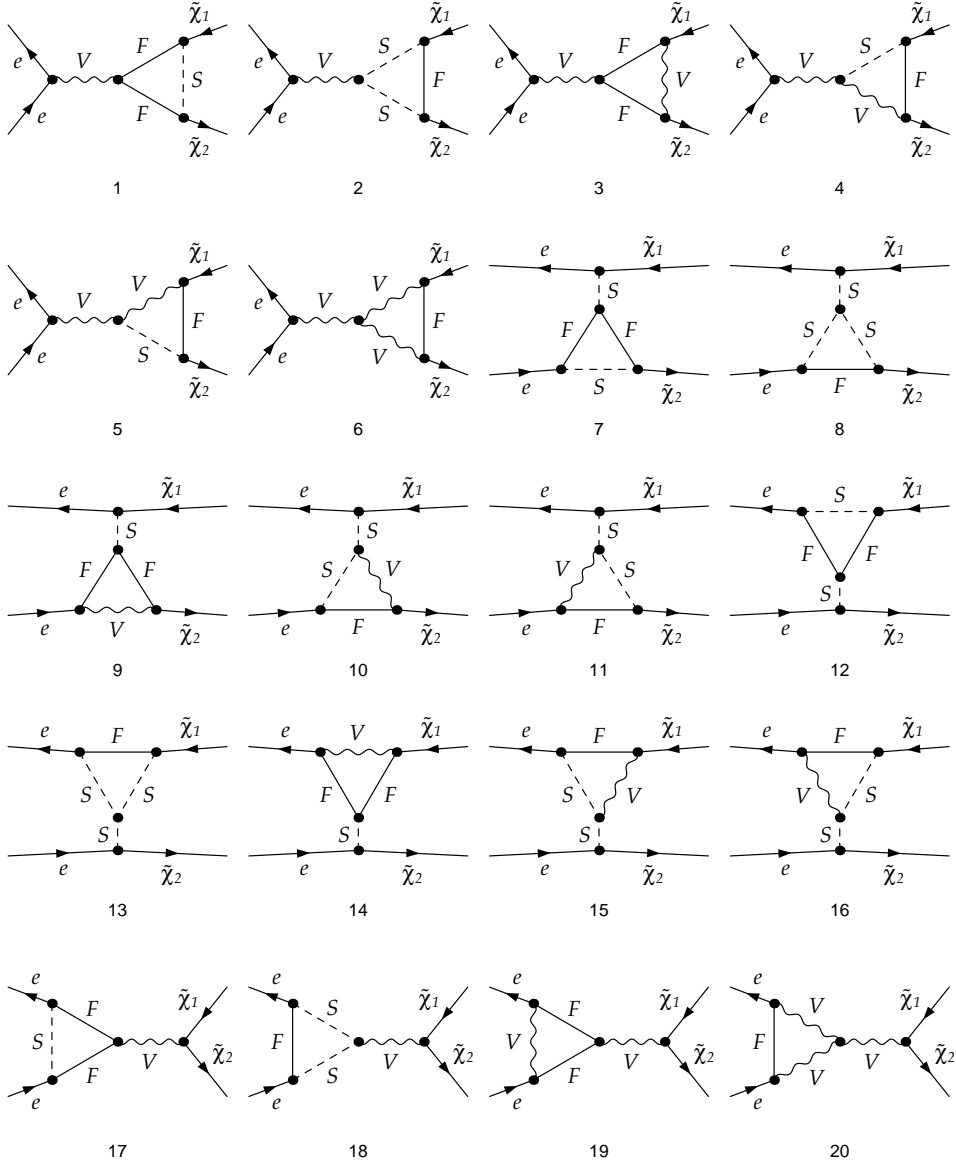}
 \end{center}
\vspace*{-0.7cm} \caption[{\bf Figure 1:}]
         {\it Generic triangle graphs contributing to
 chargino pair $\tilde{\chi}^-_2 \tilde{\chi}^+_1$ production in
         $e^+e^-$ collisions.}
\label{vert}
\end{figure}

\begin{figure}[!t]
\begin{center}

\includegraphics[scale=0.9]{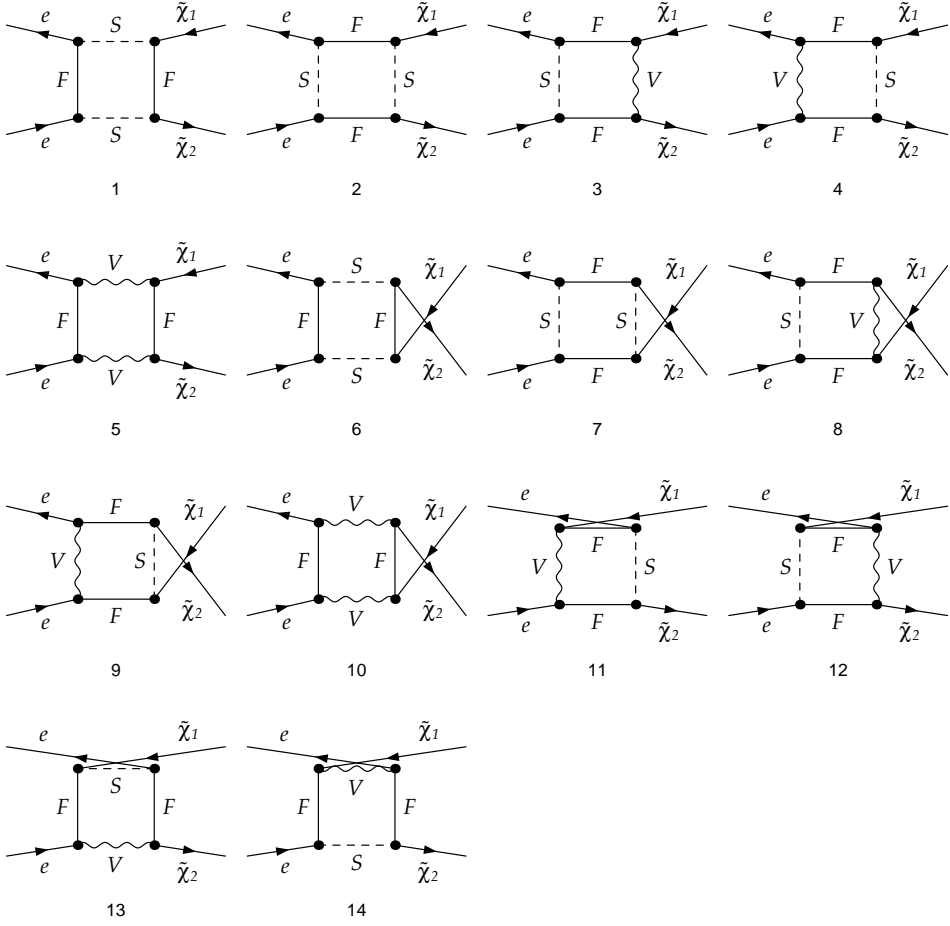}

 \end{center} \vspace*{-0.7cm}
\caption[{\bf Figure 2:}]
         {\it Generic box graphs contributing to
 chargino  pair $\tilde{\chi}^-_2 \tilde{\chi}^+_1$ production in
         $e^+e^-$ collisions.}
\label{box}
\end{figure}

Radiative corrections to the chargino pair production include the
following generic one-loop Feynman diagrams: the virtual vertex
corrections Fig.~\ref{vert}, the self-energy corrections to the
$\tilde{\nu}$, $Z$ and $\gamma$ propagators, and the box diagrams
contributions Fig.~\ref{box}. We also have to include corrections on
external chargino legs. Generation and calculation of one-loop
graphs is performed using \texttt{FeynArts~3.2} and
\texttt{FormCalc~5.2} packages \cite{feynarts}. For numerical
evaluation of loop integrals we use \texttt{LoopTools~2.2}
\cite{looptools}.

In the Ref.~\cite{Osland:2007xw} sample calculations of box diagrams
with only photon, $Z$ and $W$ boson exchanges (c.f.\ diagrams 5 and
10 in Fig.~\ref{box}) and neglecting all sfermion contributions have
been performed to demonstrate non-zero asymmetry $A_{12}$ at
one-loop. Here we present the full calculation, including all
possible contributions at the one-loop level taking into account
CP-violating phases. Full calculation of radiative corrections to
chargino pair production without CP-violating phases can be found
in~\cite{chargino-loop}.

One-loop corrected matrix element squared is given by
\begin{eqnarray}
|\mathcal{M}_{\mathrm{loop}}|^2 = |\mathcal{M}_{\mathrm{tree}}|^2 +
2 \mathrm{Re}(\mathcal{M}_{\mathrm{tree}}^*
\mathcal{M}_{\mathrm{loop}} )\, .
\end{eqnarray}
Accordingly, the one-loop CP asymmetry for the non-diagonal chargino
pair is defined as
\begin{eqnarray}
&& A_{12}=\frac{\int_{-1}^1({\rm d}\sigma^{\{12\}}_{\rm loop}-
  {\rm d}\sigma^{\{21\}}_{\rm loop}) {\rm d} \cos\theta}{\int_{-1}^1({\rm d}\sigma^{\{12\}}_{\rm tree}+
  {\rm d}\sigma^{\{21\}}_{\rm tree}){\rm d} \cos\theta}\, .
  \label{CPasym}
\end{eqnarray}
Since, as mentioned in the previous section, the CP-odd contribution
vanishes at tree level, it has to be UV-finite.
In fact one can note, that the structure of counterterms is the same
as the tree level graphs, so using the same arguments as in
Sec.~\ref{sec2} it can be shown that renormalization procedure will
not give rise to the asymmetry. Nevertheless self-energy and vertex
corrections are UV-divergent, and proper treatment of divergences is
needed. We choose to work in the dimensional reduction
scheme~\cite{siegel}, which preserves supersymmetry.

Loop diagrams with internal photon line also introduce infrared
singularities. They can be removed by adding emission of soft
photons from external charged particles. The sum of both
contributions is then IR finite, however it depends on the soft
photon cut. On the other hand soft photon emission part has the form
of tree-level amplitude multiplied by soft photon factor
\cite{denner}. Therefore, as it was explained in Sec.~\ref{sec2},
the terms arising due to soft photon bremsstrahlung do not affect
the asymmetry $A_{12}$. Similar arguments apply for hard photon
emission from external fermions.

At this point we want to make some remarks about the origin of the
CP-odd asymmetry $A_{12}$. In order to obtain a non-zero asymmetry
in the chargino production it is not enough to have CP-violating
phases in the MSSM lagrangian. In addition it requires a non-trivial
imaginary part from Feynman diagrams -- the absorptive part. Such
contributions appear when some of the intermediate state particles
in loop diagrams go on-shell. CP-odd asymmetry is generated due to
the interference between imaginary part of loop integrals and
imaginary parts of the couplings. As one can see from Eq.\
(\ref{cij_coeff}), the contributions for the production of
non-diagonal chargino pairs $\{12\}$ and $\{21\}$ differ by the
opposite sign of the imaginary part. Since the absorptive parts of
loop integrals are the same for both processes, we clearly see that
the final \textit{real} contribution to the matrix element squared
will be different in each of these final states.

\section{Numerical analysis}
To illustrate relative weights of various contributions to the CP
asymmetry, we consider two scenarios: (A) which is close to the SPS1a'
point that has been studied particularly widely~\cite{spa}; (B)~for comparison with
Ref.~\cite{Osland:2007xw}. In both
scenarios the value of the ratio of the vacuum expectation values
for Higgs fields is taken to be $\tan\beta=10$ and the parameters
defined below are low-scale parameters. \\[3mm]
In scenario (A) we take the following values for the gaugino and
higgsino mass parameters:
$$|M_1| = 100\mbox{ GeV},\quad M_2 = 200\mbox{ GeV}, \quad|\mu| = 400\mbox{ GeV}.$$
For the sfermion mass parameters we
assume
\begin{eqnarray*}
&&m_{\tilde{q}}\equiv M_{\tilde{Q}_{1,2}}=M_{\tilde{U}_{1,2}}=M_{\tilde{D}_{1,2}}=450\mbox{ GeV},\nonumber\\
&&M_{\tilde{Q}}\equiv M_{\tilde{Q}_{3}}=M_{\tilde{U}_{3}}=M_{\tilde{D}_{3}}=300\mbox{ GeV},\\
&&m_{\tilde{l}}\equiv
M_{\tilde{L}_{1,2,3}}=M_{\tilde{E}_{1,2,3}}=150\mbox{ GeV},
\end{eqnarray*}
and for the trilinear coupling
$$|A_{t}|=-A_{b}=-A_{\tau}=A=400\mbox{ GeV}.$$
Moreover, we allow non-zero phases $\Phi_\mu$ of the $\mu$
parameter, $\Phi_1$ of the bino mass parameter $M_1$ and $\Phi_t$ of
the trilinear coupling in the stop sector $A_{t}$.\\[3mm]
In scenario~(B) we take, as in Ref.~\cite{Osland:2007xw}, the
gaugino/higgsino masses
$$|M_1| = 250\mbox{ GeV}, \quad
M_2 = 200\mbox{ GeV}, \quad |\mu| = 300\mbox{ GeV}.$$ For comparison
with Ref.~\cite{Osland:2007xw} we set the universal scalar mass
$$M_{S}= M_{\tilde{Q}}= M_{\tilde{U}} =M_{\tilde{D}} =M_{\tilde{L}}=
M_{\tilde{E}}=10\mbox{ TeV},$$
so the contributions from diagrams with
exchanges of supersymmetric scalars are negligibly small. We also
investigate the $M_{S}$ dependence of the asymmetry.

In addition, the following values of the SM parameters are used:
\begin{eqnarray}
&& m_W  =  80.45\mbox{ GeV},\quad m_Z = 91.1875\mbox{ GeV},\quad
\cos\theta_W = m_W/m_Z\, ,\nonumber
\\
&& m_t = 171\mbox{ GeV},\qquad\quad\, \alpha  = 1/127.9\;.
\end{eqnarray}

The masses of relevant particles are given in Table~\ref{tab:masses}.

\begin{table*}[\hbt]
\caption[{\bf Table 2:}]{\label{tab:masses}
{\it Masses of charginos and neutralinos.}}
\begin{center}
\begin{tabular}{|l||c|c||c|c|c|c|}\hline
masses & $m_{\tilde{\chi}^\pm_1}$ & $m_{\tilde{\chi}^\pm_2}$
& $m_{\tilde{\chi}^0_1}$ & $m_{\tilde{\chi}^0_2}$ & $m_{\tilde{\chi}^0_3}$ & $m_{\tilde{\chi}^0_4}$ \\[1mm] \hline
scenario (A) & 186.7 GeV & 421.8 GeV & 97.5 GeV & 187.0 GeV & 405.8 GeV & 421.2 GeV\\[2mm]
scenario (B) & 175.6 GeV & 334.5 GeV & 172.8 GeV & 242.4 GeV & 306.5 GeV & 341.4 GeV \\
\hline
\end{tabular}
\end{center}
\end{table*}

The masses of stop squarks in scenario~(A) are
$m_{\tilde{t}_1}=204.9$~GeV and $m_{\tilde{t}_2}=438.6$~GeV. The
threshold for non-diagonal chargino pair production is $608.5$~GeV
in scenario~(A) and $510.1$~GeV in scenario~(B). Therefore for all
plots in the present analysis we take the center of mass energy
$\sqrt{s}=700$~GeV.

\begin{figure}[!t]
\begin{center}

\includegraphics[scale=0.6]{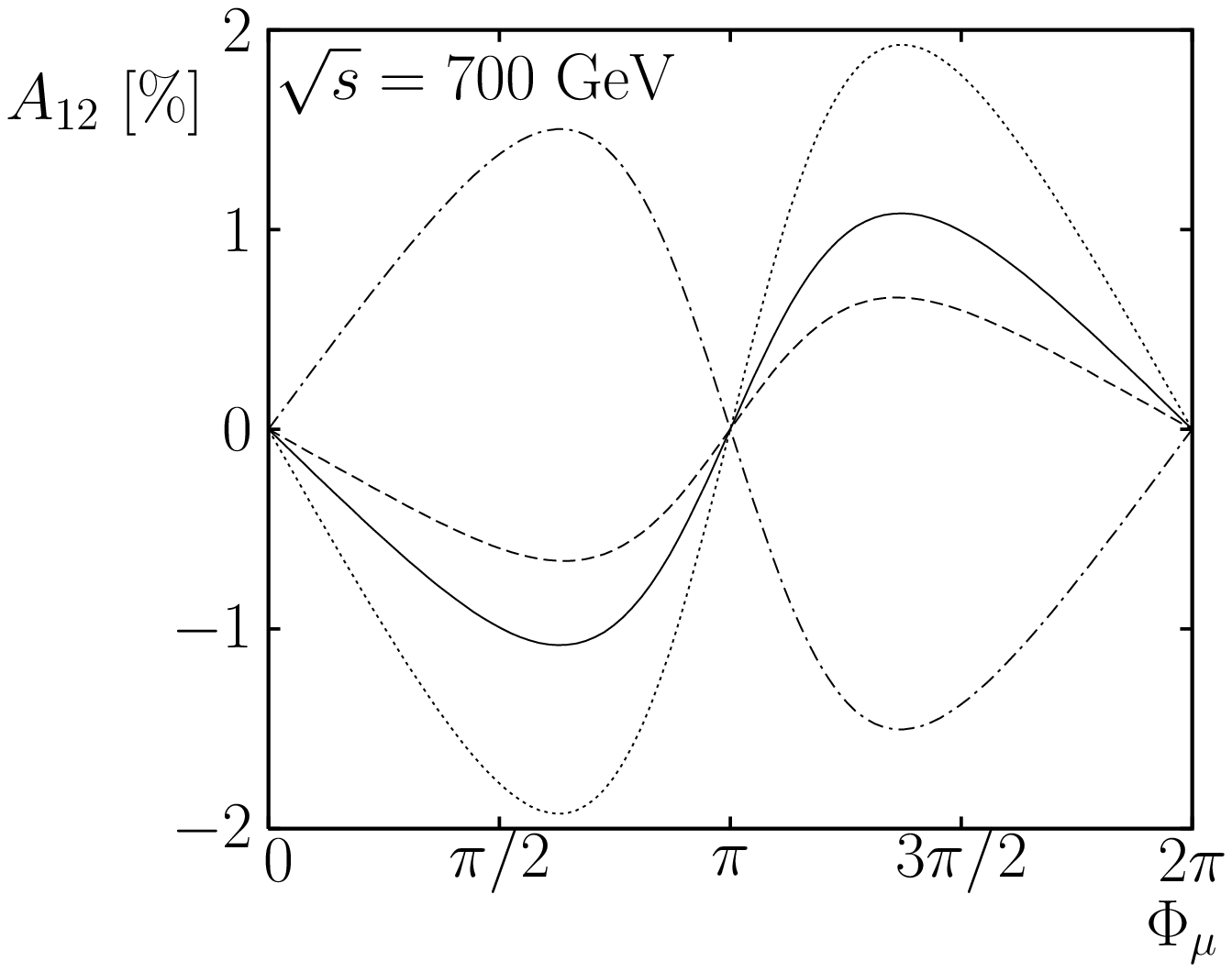}
\includegraphics[scale=0.6]{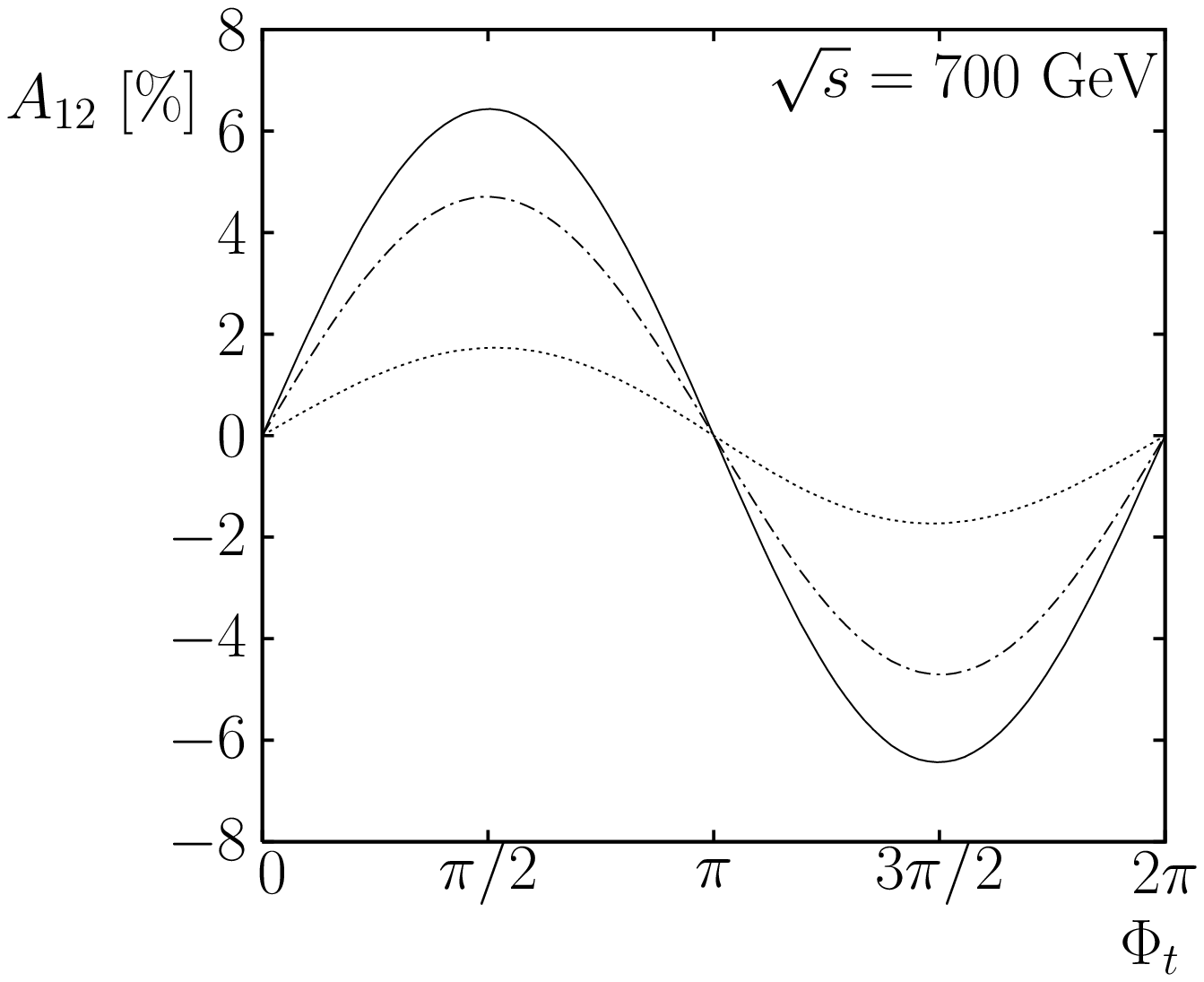}

\end{center} \vspace{-0.7cm}
\caption{\it CP asymmetry in chargino production in scenario~(A) as
a function of $\Phi_\mu$ (left) and $\Phi_t$ (right): full asymmetry
(full line) and contributions from box (dashed), vertex (dotted) and
self energy (dash-dotted) diagrams.\label{fig:asym_mu_At}}
\end{figure}

First we consider scenario~(A). The dependence of the CP asymmetry
on the phase $\Phi_\mu$ of the higgsino mass parameter $\mu$  is
shown in the left panel of Fig.~\ref{fig:asym_mu_At}. Contributions
due to box corrections, vertex corrections and self energy
corrections have been plotted in addition to the full result. The
asymmetry can reach values as large as $1\%$. Box and self-energy
diagrams can give the asymmetry of the order $1.5\%-2\%$, but since
they are of opposite signs the total asymmetry tends to be smaller.
Moreover, in this scenario the constraints from EDMs restrict the
phase $\Phi_\mu$ to be close to $n\pi$. For such values the
predicted asymmetry is very small and probably unmeasurable even at
high luminosity $e^+e^-$ linear colliders.

For the case of CP asymmetry induced by the phase $\Phi_t$ of the
trilinear coupling in the top squark sector $A_t$, the situation is
quite different, as illustrated in the right panel of
Fig.~\ref{fig:asym_mu_At}. The box diagrams do not give rise to the
CP asymmetry in this case, since there are no box diagrams with stop
exchanges. Diagrams with top squark exchanges appear only in vertex
and self energy corrections. As for vertex corrections,  only
diagrams of class 1 ($FFS = b b \tilde{t}_i$) and class 2 ($SSF =
\tilde{t}_i \tilde{t}_j b$) from Fig.~\ref{vert} contribute. The
contributions from vertex and self-energies are of the same sign and
add coherently to give the full asymmetry of the order $6\%$ -- large
enough to be measurable  in future experiments.

Note that the triangle graphs induce a coupling of the photon field
to fermions of different mass. To this coupling the diagrams of
class 1, 2, 3 and 6 with $V=\gamma$ from Fig.~\ref{vert} contribute.
These diagrams give rise to the CP asymmetry of the order $0.1\%$
both for $\Phi_\mu$ and $\Phi_t$.

\begin{figure}[!t]
\begin{center}

\includegraphics[scale=0.6]{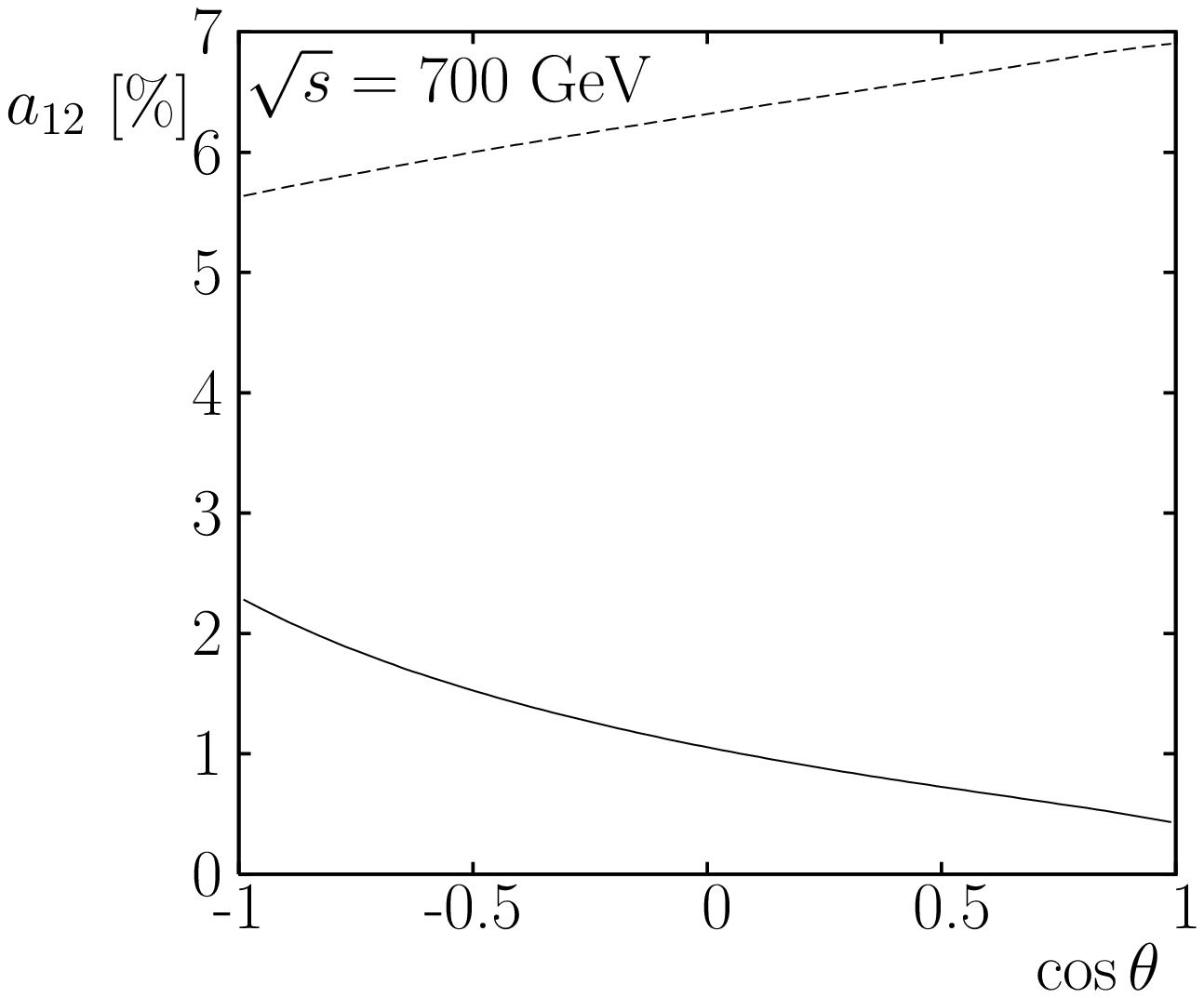}
\includegraphics[scale=0.6]{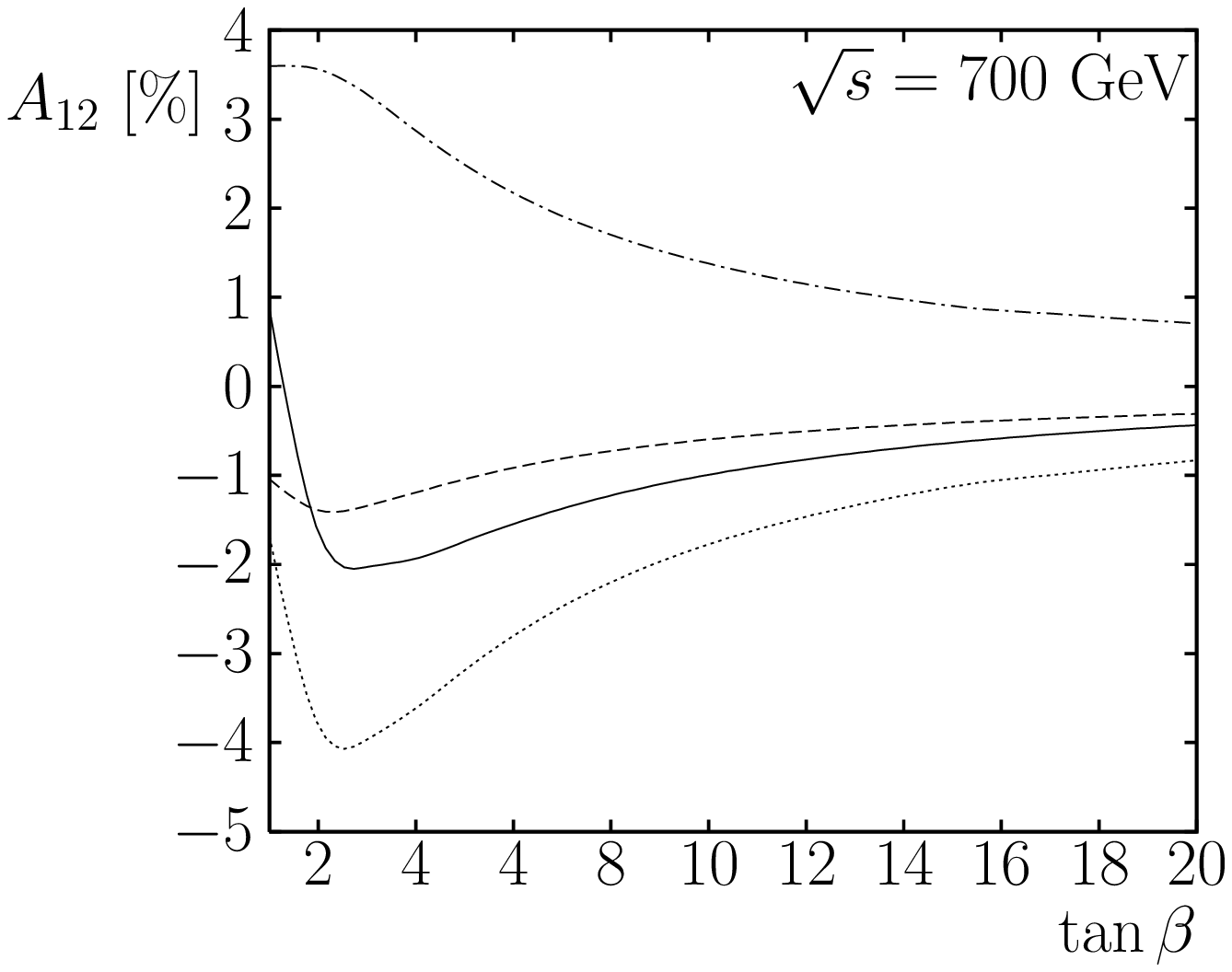}

\end{center}\vspace{-0.7cm}
\caption{{\bf Left panel: }{\it CP asymmetry (defined as in
Eq.~(\ref{CPasym}) but without integration) as a function of the
polar angle $\theta$ in scenario~(A) with $\Phi_\mu= 3\pi/2$ (full
line), and $\Phi_t= \pi/2$ (dotted line).} {\bf Right panel:} {\it
CP asymmetry Eq.~(\ref{CPasym}) as a function of $\tan\beta$ in
chargino production with other parameters as in scenario~(A): full
asymmetry (full line) and contributions from box (dashed), vertex
(dotted), self energy (dashed-dotted)
diagrams.}\label{fig:asym_diff_tbmu}}
\end{figure}

In the left panel of Fig.~\ref{fig:asym_diff_tbmu} the dependence of
the differential asymmetry $a_{12}$ (defined as in
Eq.~(\ref{CPasym}) but with integrals removed) as a function of the
production angle $\cos\theta$. For comparison we show the plots for
two choices of phases: $\Phi_\mu=3\pi/2$ (full line) and and
$\Phi_t=\pi/2$ (dotted line). Apart from the difference in
magnitude, these asymmetries have different dependence on the
$\cos\theta$: $\Phi_\mu$ asymmetry decreases with $\cos\theta$,
whereas $\Phi_t$ asymmetry increases from $5.7\%$ to $7\%$.

It is also interesting to investigate the $\tan\beta$ dependence of
the various contributions to the asymmetry. It is shown in the right
panel of Fig.~\ref{fig:asym_diff_tbmu} for the range
$\tan\beta\in[1,20]$ with $\Phi_\mu=\pi/2$, $\Phi_1=\Phi_t=0$. As
$\tan\beta$ increases from 1, the absolute values of box and vertex
contributions to the CP asymmetry increase reaching maxima around
$\tan\beta=2$ and then drop down. This behavior follows mainly from
the structure of the chargino mass matrix and consequently  the
$\tan\beta$ dependence of the chargino mixing angles and phases --
imaginary parts of coefficients $C_{12}^{R,L}$ in
Eq.~(\ref{cij_coeff}) rapidly go down for small and large values of
$\tan\beta$. For $\tan\beta=1$ the full asymmetry is close to $0$,
although again it is a result of cancelations between various
contributions.

\begin{figure}[!t]
\begin{center}

\includegraphics[scale=0.6]{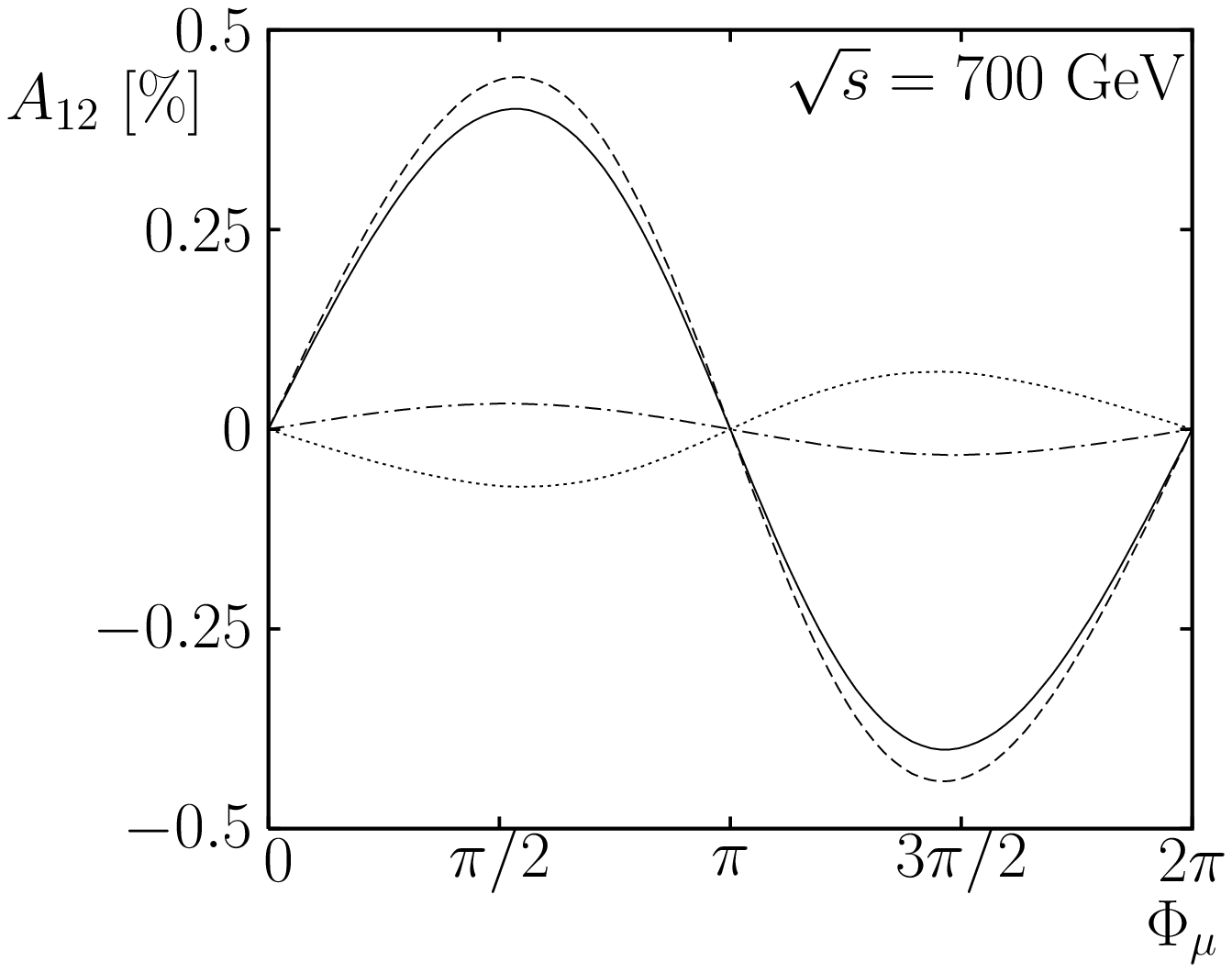}
\includegraphics[scale=0.6]{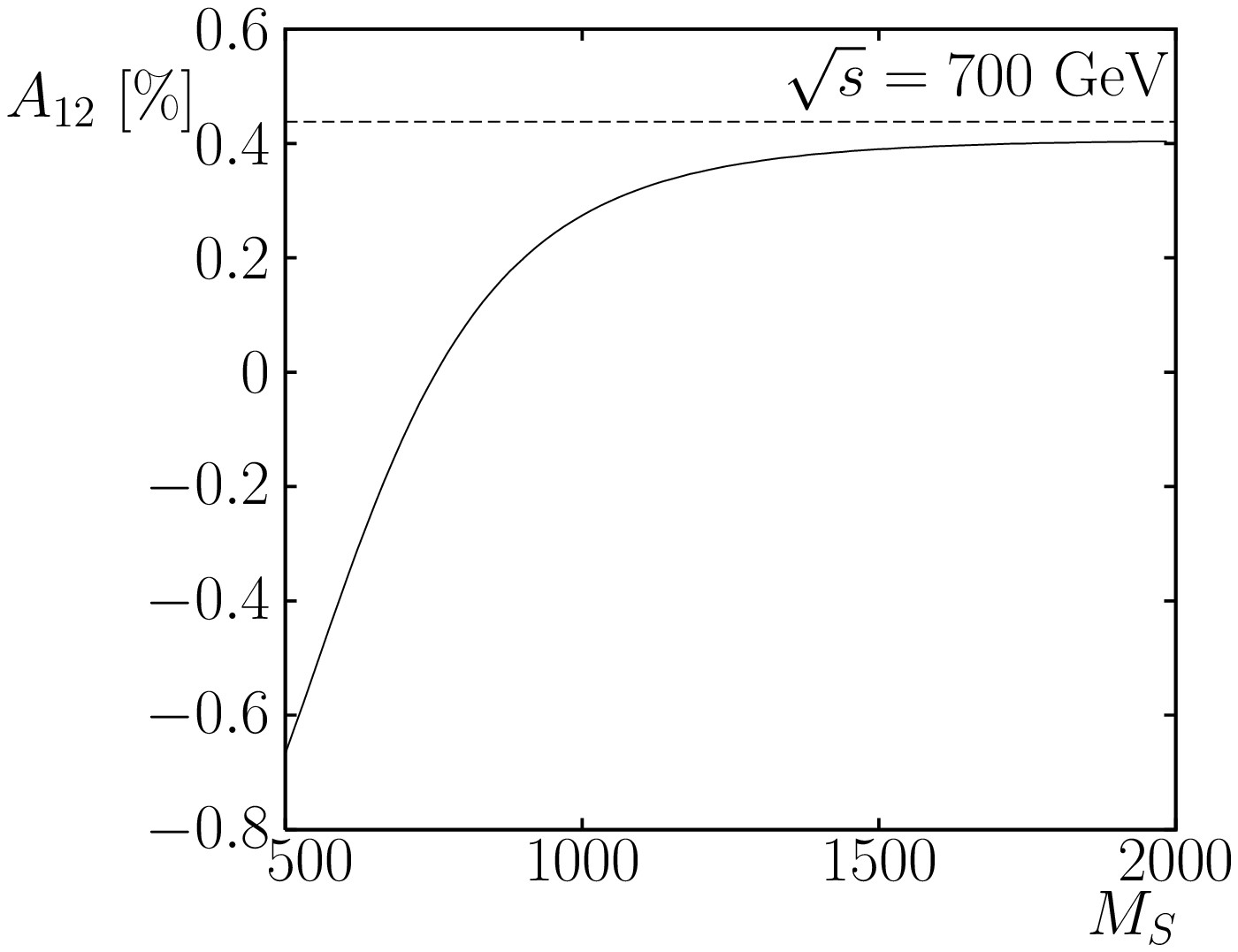}

\end{center}\vspace{-0.7cm}
\caption{{\bf Left panel: }{\it CP asymmetry in chargino production
in scenario~(B) as a function of $\Phi_\mu$ phase: full asymmetry
(full line) and contributions from box (dashed), vertex (dotted),
self energy (dashed-dotted) diagrams.} {\bf Right panel: }{\it CP
asymmetry Eq.~(\ref{CPasym}) as a function of $M_{S} =
M_{\tilde{Q}}= M_{\tilde{U}} =M_{\tilde{D}} =M_{\tilde{L}}=
M_{\tilde{E}}$. Full line is for the full result and a dotted line
is for box contributions only, as in \cite{Osland:2007xw}. Other
parameters are taken as in scenario (B) with $\Phi_\mu=
\pi/2$.}\label{fig:asym_per}}
\end{figure}

We now turn to scenario (B) as discussed in
Ref.~\cite{Osland:2007xw}. In the left panel of
Fig.~\ref{fig:asym_per} we show the full asymmetry and contributions
from box, vertex and self-energy diagrams. The asymmetry at its
maximum reaches almost $0.5\%$, and is significantly smaller than in
scenario~(A). Because sfermions are very heavy at this parameter
point, the main contribution to the asymmetry is due to box diagrams
with exchanges of vector bosons $\gamma$, $Z$, $W$, namely diagrams
of class 5 and 10 with $FFVV = e\tilde{\chi}_i \gamma Z$,
$e\tilde{\chi}_i Z \gamma$, $e \tilde{\chi}_i Z Z$, and diagram of
class 5 with $FFVV = \nu \tilde{\chi}^0_i W W$ of Fig.~\ref{box}.
Contributions from vertex and self-energy diagrams are significantly
smaller and opposite in sign and almost cancel each other.  This is
the reason why our results are consistent with results obtained
by~\cite{Osland:2007xw}. In addition, in the right panel of
Fig.~\ref{fig:asym_per} we show the dependence of the full CP
asymmetry on the  universal soft SUSY-breaking scalar mass $M_{S} =
M_{\tilde{Q}}= M_{\tilde{U}} =M_{\tilde{D}} =M_{\tilde{L}}=
M_{\tilde{E}}$ and compare it to the approximate result obtained by
Osland and Vereshagin, i.e.\ when only box contributions without
sfermion exchanges are included. With increasing $M_{S}$ the full
result approaches a constant value, which is slightly lower than the
approximate result. This small difference (which depends on $\sqrt{s}$)
is due to vertex and self-energy corrections.

We have shown the influence of the phases $\Phi_\mu$ and $\Phi_t$ on
the CP asymmetry Eq.~(\ref{CPasym}). However one can also introduce
CP violating phases for the trilinear couplings for other sfermions,
e.g.\ for bottom squarks $A_b$ and for tau sleptons $A_\tau$, as
well as for the bino mass parameter $M_1$ in the neutralino sector.
Indeed these can give rise to the CP asymmetry. However calculations
show that CP asymmetries due to these phases are typically very
small, e.g.\ for $\Phi_1$ of the order $0.1\%$, so we do not include
them here.

\section{Conclusions}
In this note we have investigated the non-diagonal chargino pair
production $e^+e^-\to\tilde{\chi}_1^\pm\tilde{\chi}_2^\mp$ at one
loop and calculated the loop-generated CP asymmetry. The CP-odd
observable can be constructed from unpolarized production cross
sections alone without the need of measuring chargino polarizations
in the final state. Our numerical analyses show that not only the
box diagrams but also the vertex and self-energy diagrams can
contribute to the CP-violation if it is induced  by the complex
higgsino mass parameter. For the case of CP-violation in the top
squark sector the box diagrams do not contribute at one loop and the
asymmetry comes entirely from vertex and self-energy diagrams. The
asymmetries can be of the order of a few percent and in principle
measurable allowing to discover the CP-violating phases via simple
event counting experiments.

\vspace{0.5cm} \noindent {\bf Acknowledgments}

\noindent We thank P.~Osland and A.~Vereshagin for useful
discussions. The authors are supported by the Polish Ministry of
Science and Higher Education Grant No.~1~P03B~108~30 and by the EU
Network MRTN-CT-2006-035505 ``Tools and Precision Calculations for
Physics Discoveries at Colliders".


\begin{thebibliography}{99}

\bibitem{genesis}A.~G.~Cohen, D.~B.~Kaplan and A.~E.~Nelson, Ann.\ Rev.\
Nucl.\ Part.\ Sci.\ {\bf 43} (1993) 27; B.~M.~Gavela, P.~Hernandez,
J.~Orloff, O.~P\`ene and C.~Quimbay, Nucl.\ Phys.B {\bf 430} (1994)
382; V.~A.~Rubakov and M.~E.~Shaposhnikov, Usp.\ Fiz.\ Nauk {\bf
166} (1996) 493.

\bibitem{susycp} S.~Pokorski, J.~Rosiek and C.~A.~Savoy, Nucl.\ Phys.B
{\bf 570} (2000) 81 [arXiv:hep-ph/9906206];
V.~Barger, T.~Falk, T.~Han, J.~Jiang, T.~Li and T.~Plehn,
Phys.\ Rev.\ D {\bf 64} (2001) 056007 [arXiv:hep-ph/0101106].

\bibitem{overcome}
  Y.~Kizukuri and N.~Oshimo,
  Phys.\ Rev.\  D {\bf 46} (1992) 3025;
  T.~Ibrahim and P.~Nath,
  Phys.\ Rev.\ D {\bf 58} (1998) 111301 [Erratum-ibid.\ D {\bf 60}
  (1998) 099902] [arXiv:hep-ph/9807501] and [arXiv:hep-ph/0107325];
  T.~Ibrahim and P.~Nath,
  Phys.\ Rev.\ D {\bf 61}, 093004 (2000) [arXiv:hep-ph/9910553];
  M.~Brhlik, G.~J.~Good and G.~L.~Kane,
  Phys.\ Rev.\  D {\bf 59} (1999) 115004
  [arXiv:hep-ph/9810457];
  S.~Abel, S.~Khalil and O.~Lebedev,
  Nucl.\ Phys.\  B {\bf 606} (2001) 151
  [arXiv:hep-ph/0103320];
  R.~Arnowitt, B.~Dutta and Y.~Santoso, Phys.\ Rev.\ D {\bf 64} (2001)
  113010
  [arXiv:hep-ph/0106089];
  T.~Ibrahim and P.~Nath,
  arXiv:0705.2008 [hep-ph].

\bibitem{Kizukuri:1993vh}
  Y.~Kizukuri and N.~Oshimo,
  arXiv:hep-ph/9310224.

\bibitem{spin}
A.~Bartl, H.~Fraas, O.~Kittel and W.~Majerotto,
  Phys.\ Rev.\  D {\bf 69} (2004) 035007
  [arXiv:hep-ph/0308141];
A.~Bartl, T.~Kernreiter and O.~Kittel,
  Phys.\ Lett.\  B {\bf 578} (2004) 341
  [arXiv:hep-ph/0309340];
A.~Bartl, K.~Hohenwarter-Sodek, T.~Kernreiter and H.~Rud,
  Eur.\ Phys.\ J.\  C {\bf 36} (2004) 515
  [arXiv:hep-ph/0403265];
A.~Bartl, H.~Fraas, S.~Hesselbach, K.~Hohenwarter-Sodek, T.~Kernreiter and G.~A.~Moortgat-Pick,
  JHEP {\bf 0601} (2006) 170
  [arXiv:hep-ph/0510029];
A.~Bartl, K.~Hohenwarter-Sodek, T.~Kernreiter and O.~Kittel,
  arXiv:0706.3822 [hep-ph];
S.~Y.~Choi, M.~Drees and J.~Song,
  JHEP {\bf 0609} (2006) 064
  [arXiv:hep-ph/0602131].

\bibitem{Kittel:2004kd}
  A.~Bartl, H.~Fraas, O.~Kittel and W.~Majerotto,
  Phys.\ Lett.\  B {\bf 598} (2004) 76
  [arXiv:hep-ph/0406309];
  O.~Kittel, A.~Bartl, H.~Fraas and W.~Majerotto,
  Phys.\ Rev.\  D {\bf 70} (2004) 115005
  [arXiv:hep-ph/0410054].

\bibitem{Choi:2001ww}
  S.~Y.~Choi, J.~Kalinowski, G.~A.~Moortgat-Pick and P.~M.~Zerwas,
  Eur.\ Phys.\ J.\  C {\bf 22} (2001) 563
  [arXiv:hep-ph/0108117],
  Addendum-ibid.\  C {\bf 23} (2002) 769 [arXiv:hep-ph/0202039];
J.~Kalinowski,
  Acta Phys.\ Polon.\  B {\bf 34} (2003) 3441
  [arXiv:hep-ph/0306272].



\bibitem{Choi:2003hm}
  S.~Y.~Choi,
  Phys.\ Rev.\  D {\bf 69} (2004) 096003
  [arXiv:hep-ph/0308060];
  S.~Y.~Choi, B.~C.~Chung, J.~Kalinowski, Y.~G.~Kim and K.~Rolbiecki,
  Eur.\ Phys.\ J.\  C {\bf 46} (2006) 511
  [arXiv:hep-ph/0504122].

\bibitem{Osland:2007xw}
  P.~Osland and A.~Vereshagin,
  Phys.\ Rev.\  D {\bf 76} (2007) 036001
  [arXiv:0704.2165 [hep-ph]].

\bibitem{Choi:charg}
  S.~Y.~Choi, A.~Djouadi, H.~K.~Dreiner, J.~Kalinowski and P.~M.~Zerwas,
  Eur.\ Phys.\ J.\  C {\bf 7} (1999) 123
  [arXiv:hep-ph/9806279];
S.~Y.~Choi, A.~Djouadi, H.~S.~Song and P.~M.~Zerwas,
  Eur.\ Phys.\ J.\  C {\bf 8} (1999) 669
  [arXiv:hep-ph/9812236];
S.~Y.~Choi, M.~Guchait, J.~Kalinowski and P.~M.~Zerwas,
  Phys.\ Lett.\  B {\bf 479} (2000) 235
  [arXiv:hep-ph/0001175];
S.~Y.~Choi, A.~Djouadi, M.~Guchait, J.~Kalinowski, H.~S.~Song and P.~M.~Zerwas,
  Eur.\ Phys.\ J.\  C {\bf 14} (2000) 535
  [arXiv:hep-ph/0002033].

\bibitem{futurecoll} TESLA Technical Design Report, Part: III Physics at
an $e^+e^-$ Linear Collider, {\it eds.}\ R.-D.~Heuer, D.~Miller, F.~Richard
and P.~Zerwas, DESY 2001-011, arXiv:hep-ph/0106315;
T.~Abe {\it et al.}  [American Linear Collider Working Group Collaboration],
``Linear collider physics resource book for Snowmass 2001. 2: Higgs
and  supersymmetry studies,''
in {\it Proc. of the APS/DPF/DPB Summer Study on the Future of
  Particle Physics (Snowmass 2001) } ed. N.~Graf,
arXiv:hep-ex/0106056;
K. Abe {\it et al.}, ILC Roadmap Report, presented at the ACFA LC
Symposium,  Tsukuba, Japan 2003, http://lcdev.kek.jp/RMdraft/ .

\bibitem{feynarts}
J.~K\"{u}blbeck, M.~Bohm and A.~Denner,
Comput.\ Phys.\ Commun.\ {\bf 60} (1990) 165; T.~Hahn,
Comput.\ Phys.\ Commun.\ {\bf 140} (2001) 418
[arXiv:hep-ph/0012260v2];
T.~Hahn and M.~Perez-Victoria,
Comput.\ Phys.\ Commun.\ {\bf 118} (1999) 153
[arXiv:hep-ph/9807565];
T.~Hahn and C.~Schappacher,
Comput.\ Phys.\ Commun.\ {\bf 143} (2002) 54 [arXiv:hep-ph/0105349].

\bibitem{looptools}
G.~J.~van~Oldenborgh,
Comput.\ Phys.\ Commun.\ {\bf 66} (1991) 1; T.~Hahn,
Acta.\ Phys.\ Pol.\ B {\bf 30} (1999) 3469 [arXiv:hep-ph/9910227].

\bibitem{chargino-loop}
  T.~Fritzsche and W.~Hollik,
  Nucl.\ Phys.\ Proc.\ Suppl.\  {\bf 135} (2004) 102
  [arXiv:hep-ph/0407095];
  W.~Oller, H.~Eberl and W.~Majerotto,
  Phys.\ Rev.\  D {\bf 71} (2005) 115002
  [arXiv:hep-ph/0504109].

\bibitem{siegel}
  W.~Siegel,
  Phys.\ Lett.\ {\bf 84} (1979) 193;
  D.~M.~Capper, D.~R.~T.~Jones and P.~van Nieuwenhuizen,
  Nucl.\ Phys.\  B {\bf 167} (1980) 479;
  D.~St\"{o}ckinger,
  JHEP {\bf 0503} (2005) 076
  [arXiv:hep-ph/0503129].

\bibitem{denner}
  A.~Denner,
  Fortsch.\ Phys.\  {\bf 41} (1993) 307
  [arXiv:0709.1075 [hep-ph]].

\bibitem{spa}
  J.~A.~Aguilar-Saavedra {\it et al.},
  Eur.\ Phys.\ J.\ C {\bf 46} (2006) 43 [hep-ph/0511344].
\end{thebibliography}
\end{document}